\documentclass{jpsj2}
\def\la{\mathrel{\mathpalette\fun <}}
\def\ga{\mathrel{\mathpalette\fun >}}
\def\fun#1#2{\lower3.6pt\vbox{\baselineskip0pt\lineskip.9pt
  \ialign{$\mathsurround=0pt#1\hfil##\hfil$\crcr#2\crcr\sim\crcr}}}

\title{Properties of Nuclear and Coulomb Breakup of $^8$B}

\author{
Kazuyuki \textsc{Ogata}$^{1}$\thanks{E-mail: ogata@phys.kyushu-u.ac.jp},
Takuma \textsc{Matsumoto}$^{2}$\thanks{Present address: Meme Media Laboratory, Hokkaido University, Sapporo 060-8628, Japan},
Yasunori \textsc{Iseri}$^{3}$,
and Masanobu \textsc{Yahiro}$^{1}$}

\inst{$^{1}$Department of Physics, Kyushu University,
6-10-1 Hakozaki, Fukuoka 812-8581, Japan \\
$^{2}$RIKEN Nishina Center, 2-1 Hirosawa, Wako 351-0198, Japan \\
$^{3}$Department of Physics, Chiba-Keizai College,
4-3-30 Todoroki-cho, Inage, Chiba 263-0021, Japan}

\abst{The dependence of breakup cross sections of $^8$B
at 65 MeV/nucleon on the
target mass number $A_{\rm T}$ is investigated by means
of the continuum-discretized coupled-channels method (CDCC) with
more reliable distorting potentials than those in the preceding study.
The $A_{\rm T}^{1/3}$ scaling law of the nuclear breakup cross section
is found to be satisfied only in the middle $A_{\rm T}$ region of
$40 \la A_{\rm T} \la 150$.
The interference between nuclear and Coulomb breakup amplitudes
vanishes in very forward angle scattering,
independently of the target nucleus.
The truncation of the relative energy between the $p$ and $^7$Be
fragments slightly reduces the contribution of nuclear breakup
at very forward angles, while the angular region in which
the first-order perturbation theory works well does not
change essentially.
}

\kword{CDCC, breakup reaction, $^8$B, nuclear-Coulomb interference,
multistep processes}

\begin{document}
\maketitle

\section{Introduction}

The properties of unstable nuclei are one of the most important
subjects in nuclear physics. The breakup reactions of such short-lived
nuclei provide us with much information on their static and dynamical
features. The responses to electromagnetic fields of unstable nuclei,
which have been intensively studied in particular,
are expected to be obtained using breakup reactions induced by a heavy target
nucleus such as $^{208}$Pb because of its dominant Coulomb field
compared with the nuclear field. Recently, studies of two-neutron halo
nuclei, e.g., $^6$He and $^{11}$Li, based on this conjecture have been
carried out, and B(E1) strengths and spectroscopic
factors were evaluated.\cite{6He,11Li} \
The extraction of {\lq\lq}pure'' responses to
electromagnetic fields, however, requires an accurate description
of the possible nuclear breakup, multistep transitions, and
the interference between nuclear and Coulomb breakup amplitudes.
The systematic analysis \cite{Hussein}
of the breakup reactions of $^7$Be, $^8$B, and $^{11}$Be
with the continuum-discretized coupled-channels method
(CDCC) \cite{CDCC-review}
showed that the nuclear breakup cross section $\sigma_{\rm N}$ was scaled
as $A_{\rm T}^{1/3}$, where $A_{\rm T}$ is the target mass number.
If the scaling is true, one can evaluate $\sigma_{\rm N}$ induced by
a heavy target such as $^{208}$Pb,
from a measured breakup cross section by
a light target such as $^{12}$C; the latter is expected to contain
only nuclear breakup contribution.
Thus, there is a possibility that one can
remove the {\lq\lq}contamination''
due to nuclear breakup
from the measured breakup cross section by a heavy target, which
is generally due to both nuclear and Coulomb breakup.
It was also shown,\cite{Hussein} however,
that one could not be free of nuclear-Coulomb (N-C) interference,
even if only events corresponding to forward-angle scattering,
in which the scattering angle of the center of mass (c.m.)
of the projectile is small, are selected. An accurate analysis
of the breakup reactions of unstable nuclei with CDCC, therefore, is
necessary to draw a quantitative conclusion on B(E1) values and
spectroscopic factors. A four-body CDCC \cite{fb-CDCC}
analysis of breakup reactions of $^6$He and $^{11}$Li
is particularly interesting and important.

Before discussing the four-body system, in this study we
reinvestigate the breakup reactions of the $^8$B nucleus at
65 MeV/nucleon by several target nuclei ${\rm A}$, i.e.,
$^{12}$C, $^{16}$O, $^{40}$Ca, $^{58}$Ni, $^{90}$Zr, $^{152}$Sm,
and $^{208}$Pb,
with CDCC based on a $^7$Be$+p+$A three-body model.
The main purpose of this study is to show some results
using more realistic $p$-A and $^7$Be-A optical potentials
than those in the preceding study.\cite{Hussein}
The $A_{\rm T}^{1/3}$ scaling law of $\sigma_{\rm N}$ and N-C
interference are investigated.
As an interesting point not discussed in ref.~\citen{Hussein},
the dependence of N-C interference
on the cut off relative energy between
the $^7$Be and $p$ fragments after breakup is shown here.

This paper is constructed as follows. In \S\ref{num}, we describe
details of numerical inputs of CDCC. In \S\ref{res},
the results of the present calculation
are shown, and the $A_{\rm T}^{1/3}$
scaling law of $\sigma_{\rm N}$ and N-C interference
are discussed.
Finally, we give a summary in \S\ref{sum}.

\section{Numerical Calculation}
\label{num}

We follow the formulation of breakup cross sections with CDCC
described in ref.~\citen{Ogata} except that we disregard the
channel-spin of $^8$B.
We take the $^8$B wave function in
ref.~\citen{EB} for the p-wave, with neglecting the spin-orbit
potential; the depth of the central part is adjusted to reproduce
the proton separation energy $S_p=137$ keV. For the s-wave,
the $p$-$^7$Be potential of Barker \cite{Barker}
that reproduces the scattering length \cite{Angulo} for the channel
spin $S=2$ component is adopted. As for the d- and f-state
potentials, we use the parameters of ref.~\citen{Typel}.
We take 20, 10, 10, and 5 discretized continuum states
for the s-, p-, d-, and f-waves, respectively.
The maximum wave number (radius)
between $p$ and $^7$Be is set to be 0.66 fm$^{-1}$ (200 fm).
When only nuclear breakup is included, the scattering waves
between $^8$B and the target are integrated up to
$R_{\rm max}=100$ fm.
The values of $R_{\rm max}$ for the calculation of
nuclear and Coulomb breakup are 210, 210, 250, 330, 450,
670, and 880 fm for the $^{12}$C, $^{16}$O, $^{40}$Ca, $^{58}$Ni,
$^{90}$Zr, $^{152}$Sm, and $^{208}$Pb targets, respectively.
The maximum orbital angular momentum $L_{\rm max}$ between $^8$B and
$A$ is evaluated as $L_{\rm max}=K_{\rm A} R_{\rm max}$, where
$K_{\rm A}$ is the $^8$B-${\rm A}$ relative wave number in the
incident channel.
For the partial waves with $L>1000$, the eikonal-CDCC method
(E-CDCC) \cite{Ogata,Ogata2} is used to obtain the $S$-matrix elements.

We adopt the global optical potential based on Dirac phenomenology
(the EDAD1 parameter) \cite{Hama}
for the distorting potential between $p$ and A.
As for the $^7$Be-A potential $U(R_{7})$,
we slightly modify the $^7$Li global optical potential of
Cook \cite{Cook}, which is appropriate at approximately 10 MeV/nucleon, as
\begin{eqnarray}
U(R_{7})&=&
-74.03\left\{1+\exp\left[\frac{R_{7}-1.325A_{\rm T}^{1/3}}{0.853}
\right]\right\}^{-1}
\nonumber \\
&&
-iW(A_{\rm T})\left\{1+\exp\left[\frac{R_{7}-1.640A_{\rm T}^{1/3}}{0.809}
\right]\right\}^{-1},
\nonumber \\
W(A_{\rm T})&=&31.47-0.160A_{\rm T}+0.00045 A_{\rm T}^2.
\nonumber
\end{eqnarray}
Note that the diffuseness parameters are not changed, and the
$A_{\rm T}$ dependence appears only in the radial parameters and
the depth of the imaginary part as in the original potential.\cite{Cook}
This potential is found to reproduce quite well the elastic cross sections
of $^7$Li-$^{12}$C at 67.8 MeV/nucleon \cite{Gil} and of $^7$Be-$^{208}$Pb
at 60.8 MeV/nucleon.\cite{Bishop}

\section{Results and Discussion}
\label{res}

First we examine the $A_{\rm T}^{1/3}$ scaling law of $\sigma_{\rm N}$,
which is based on the following two assumptions \cite{Hussein}:
1) the partial breakup cross section
$\sigma_{\rm N} (L)$ is localized around the grazing orbital angular
momentum, i.e., $L_{\rm g} \equiv K_{\rm A} (R_{\rm P} + R_{\rm T})$
with $R_{\rm P}$ ($R_{\rm T}$) being the radius of the projectile (target);
and
2) the transmission coefficient at $L_{\rm g}$, $T_{L_{\rm g}}$,
is independent of the target nuclei.
%
\begin{figure}[htbp]
\begin{center}
\includegraphics[width=0.45\textwidth,clip]{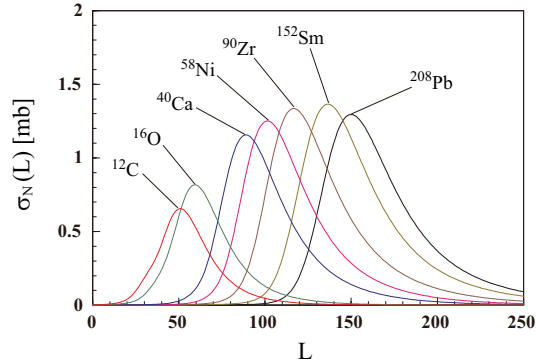}
\end{center}
\caption{
(Color online)
The partial cross sections for nuclear breakup of $^8$B by the
seven target nuclei concerned.
}
\label{fig1}
\end{figure}
In Fig.~\ref{fig1} we show $\sigma_{\rm N} (L)$ for the seven target nuclei
calculated with CDCC, with neglecting Coulomb breakup.
The values of $L_{\rm g} / K_{\rm A}$ and
\[T_{L_{\rm g}} =
\frac{\sigma_{\rm N} (L_{\rm g}) K_{\rm A}^2}{\pi (2 L_{\rm g}+1)}
\]
extracted are shown in the second and fourth rows in Table \ref{tab1},
respectively.
%
\begin{table}[htbp]
\caption{
$L_{\rm g} / K_{\rm A}$ and $T_{L_{\rm g}}$ extracted from
partial cross sections shown in Fig.~\ref{fig1}. See text for details.
}
\begin{tabular}{cccccccc}
\hline
$A$\;&
\;$^{12}$C\;   &   \;$^{16}$O\;   &   \;$^{40}$Ca\;   &   \;$^{58}$Ni\;   &
\;$^{90}$Zr\;  &   \;$^{152}$Sm\; &   \;$^{208}$Pb\;
\\ \hline
$L_{\rm g} / K_{\rm A}$ [fm]\;  &
\;6.02\; & \;6.27\; & \;7.54\; & \;8.15\; & \;9.00\; & \;10.1\; & \;11.0\; \\
$1.4A_{\rm T}^{1/3} +2.7$ [fm]\;  &
\;5.91\; & \;6.23\; & \;7.49\; & \;8.12\; & \;8.97\; & \;10.2\; & \;11.0\; \\
$T_{L_{\rm g}}$\;  &
\;0.14\; & \;0.19\; & \;0.29\; & \;0.30\; & \;0.31\; & \;0.28\; & \;0.25\;
\\ \hline
\end{tabular}
\label{tab1}
\end{table}
One sees that $L_{\rm g} / K_{\rm A}$
is scaled as $A_{\rm T}^{1/3}$, i.e.,
$1.4A_{\rm T}^{1/3} +2.7$
(the third row) reproduces $L_{\rm g} / K_{\rm A}$ very well.
This fact justifies  the interpretation of the extracted $L_{\rm g}/K_{\rm A}$
as $R_{\rm P} + R_{\rm T}$. Namely, the assumption 1) above is shown to be
valid. The corresponding $T_{L_{\rm g}}$, however, shows
a clear target dependence. In the middle $A_{\rm T}$ region of
$40 \le A_{\rm T} \le 152$, $T_{L_{\rm g}}$ is almost constant, while
$T_{L_{\rm g}}$ significantly decreases for $A_{\rm T} \le 16$ and $A_{\rm T} = 208$.
This indicates that, in the reactions with a very light or very heavy target,
the nuclear absorption at $R_{\rm P} + R_{\rm T}$ due to
the imaginary parts of the $p$-A and $^7$Be-A optical potentials
is important compared with that in the reactions with middle-heavy targets.

Thus, the $A_{\rm T}^{1/3}$ scaling law of $\sigma_{\rm N}$
holds in the $40 \la A_{\rm T} \la 152$ region.
%
\begin{figure}[htbp]
\begin{center}
\includegraphics[width=0.45\textwidth,clip]{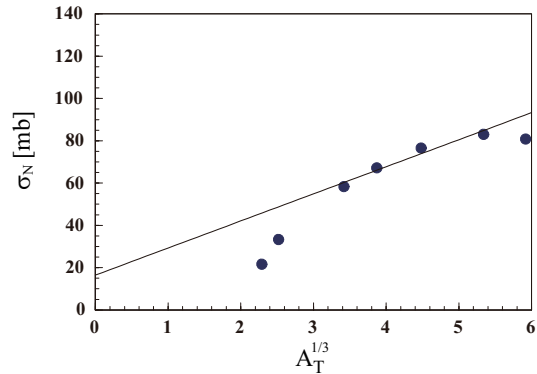}
\end{center}
\caption{
Nuclear breakup cross section $\sigma_{\rm N}$
plotted against $A_{\rm T}^{1/3}$.
The solid line shows the scaled result
$\bar{\sigma}_{\rm N}(A_{\rm T}^{1/3})
= 12.8 A_{\rm T}^{1/3} + 16.5$, obtained by fitting to
the $\sigma_{\rm N}$ for the $^{40}$Ca, $^{58}$Ni, $^{90}$Zr, and $^{152}$Sm
targets.
}
\label{fig2}
\end{figure}
In Fig.~\ref{fig2} we show the comparison between $\sigma_{\rm N}$
and the scaling formula
\[
\bar{\sigma}_{\rm N}(A_{\rm T}^{1/3}) = 12.8 A_{\rm T}^{1/3} + 16.5,
\]
which is determined to reproduce the $\sigma_{\rm N}$
for $A_{\rm T}=40$, 58, 90, and 152.
Note that whether
the $A_{\rm T}^{1/3}$ scaling law is satisfied or not depends on
the properties of the distorting potentials used in the CDCC calculation,
which dictates the actual $A_{\rm T}$ dependence of $T_{L_{\rm g}}$.
In order to
perform the systematic analysis of the breakup reactions of $^{11}$Be,
$^{7}$Be, $^{6}$He, $^{11}$Li, and others with three-body or four-body
CDCC, therefore, reliable optical potentials between the constituents of the
projectile and the targets are necessary.

\begin{figure}[htbp]
\begin{center}
\includegraphics[width=0.45\textwidth,clip]{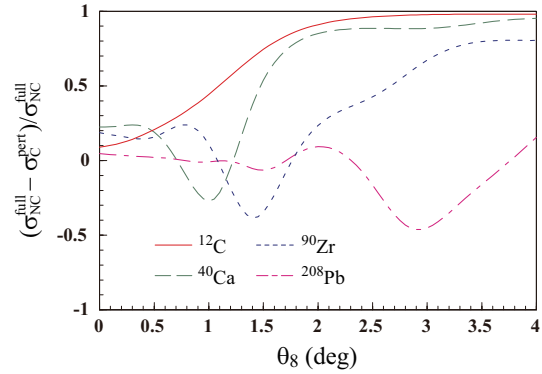}
\end{center}
\caption{
(Color online)
Relative difference between the breakup cross sections
calculated with full CDCC and first-order perturbative
CDCC, as a function of $\theta_8$. The solid, dashed, dotted, and
dash-dotted lines correspond to the breakup by
$^{12}$C, $^{40}$Ca, $^{90}$Zr, and $^{152}$Sm, respectively.
}
\label{fig3}
\end{figure}

Next, we show in Fig.~\ref{fig3} the deviation (relative difference) of the
breakup cross section $\sigma_{\rm C}^{\rm pert}$, calculated with
first-order perturbative CDCC, from the cross section
$\sigma_{\rm NC}^{\rm full}$ calculated with full CDCC.
In the former, one-step transitions due to monopole and
dipole Coulomb interactions are only included,
while in the latter all the processes due to nuclear and Coulomb
interactions are included.
The cross sections are integrated over the relative
energy $\epsilon$ between the $p$ and $^7$Be fragments up to
10 MeV.
%
The solid, dashed, dotted, and dash-dotted lines in Fig.~\ref{fig3}
indicate the
results for the $^{12}$C, $^{40}$Ca, $^{90}$Zr, and $^{208}$Pb
targets, respectively.
The horizontal axis is the c.m. scattering angle $\theta_8$ of $^8$B.
As expected, the difference is very large except at
forward angles
because of the dominant contribution of nuclear breakup.
It should be noted, however, that
even at very forward angles $\theta_8 \sim 0^\circ$
the difference is somewhat large, i.e., about 10--20\%,
for the breakup by $^{12}$C, $^{40}$Ca, and $^{90}$Zr targets.
On the other hand, for the breakup by $^{208}$Pb,
the difference is less than 10\% for $\theta_8 \le 2.4^\circ$,
which justifies in part the use of the first-order
perturbation theory to extract {\lq\lq}pure'' responses of $^8$B
to electromagnetic fields in this reaction.
The possible 10\% error of such simplified analysis,
however, should be noted.

%
\begin{figure}[htbp]
\begin{center}
\includegraphics[width=0.45\textwidth,clip]{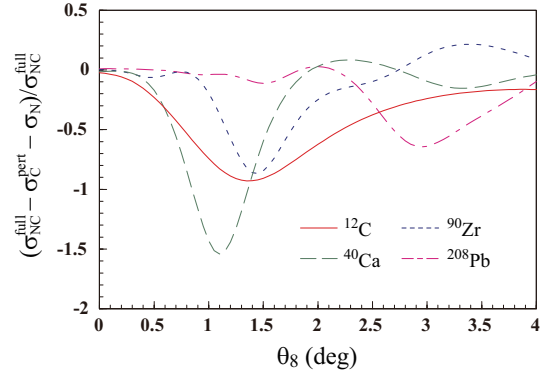}
\end{center}
\caption{
(Color online)
Same as in Fig.~\ref{fig3} but for the relative difference between
the breakup cross section calculated with full CDCC and that
with first-order perturbative CDCC added by
the nuclear breakup cross section.
}
\label{fig4}
\end{figure}
Figure \ref{fig4} shows the relative difference between
$\sigma_{\rm NC}^{\rm full}$ and $\sigma_{\rm C}^{\rm pert}+\sigma_{\rm N}$.
It is found that multistep processes due to Coulomb breakup are
negligible in the present case. Thus,
$\sigma_{\rm NC}^{\rm full}-(\sigma_{\rm C}^{\rm pert}+\sigma_{\rm N})$
shows the importance of N-C interference.
One sees from Fig.~\ref{fig4} that, except for $^{208}$Pb,
the interference is crucially important for
$1.0^\circ \la \theta_8 \la 2.0^\circ$. This difference is
closely related to the value shown in Fig.~\ref{fig3}
in ref.~\citen{Hussein},
in which it was concluded that the contribution of the interference
tended to vanish only for $^8$B breakup by $^{208}$Pb.
Our present calculation shows, however, that the contribution
of N-C interference almost vanishes at $\theta_8 \sim 0^\circ$
for all target nuclei.
%
\begin{table}[htbp]
\caption{
Values of $\theta_{\rm cr}$ and $b_{\rm cr}$.
See text for details.
}
\begin{tabular}{cccccccc}
\hline
$A$\;&
\;$^{12}$C\;   &   \;$^{16}$O\;   &   \;$^{40}$Ca\;   &   \;$^{58}$Ni\;   &
\;$^{90}$Zr\;  &   \;$^{152}$Sm\; &   \;$^{208}$Pb\;
\\ \hline
$\theta_{\rm cr}$ [deg]\;  &
\;0.30\; & \;0.30\; & \;0.40\; & \;0.60\; & \;0.90\; & \;1.65\; & \;2.30\; \\
$b_{\rm cr}$ [fm]\;  &
\;26.5\; & \;31.8\; & \;47.6\; & \;42.1\; & \;38.4\; & \;31.4\; & \;29.4\;
\\ \hline
\end{tabular}
\label{tab2}
\end{table}
Table \ref{tab2}
shows the maximum scattering angle $\theta_{\rm cr}$ of the region
in which the contribution of N-C interference is less than 10\%,
together with the corresponding impact parameters $b_{\rm cr}$,
for the seven reactions concerned.

\begin{figure}[htbp]
\begin{center}
\includegraphics[width=0.45\textwidth,clip]{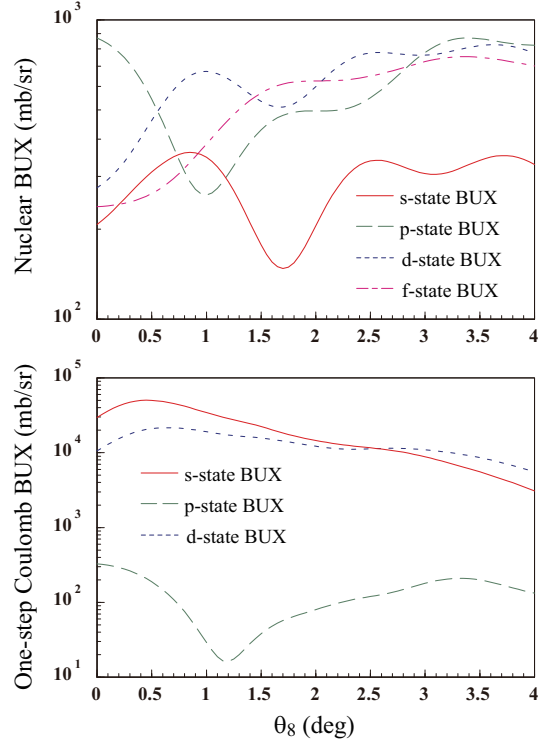}
\end{center}
\caption{
(Color online)
In the upper panel, the nuclear breakup cross section (BUX) by $^{208}$Pb
is decomposed into s- (solid line),
p- (dashed line), d- (dotted line),
and f-state (dash-dotted line) components.
The lower panel corresponds to the one-step Coulomb breakup
cross section calculated by first-order perturbative CDCC.
}
\label{fig5}
\end{figure}

The N-C vanishment at $\theta_8 \sim 0^\circ$ can be explained as follows.
In the upper panel of Fig.~\ref{fig5}, we show the
s- (solid line), p- (dashed line), d- (dotted line),
and f-state (dash-dotted line) breakup components of the
nuclear breakup cross section of $^8$B by $^{208}$Pb.
One sees that the p-state breakup is dominant at very forward angles.
This is because the monopole nuclear coupling
potentials are significantly larger than those with other
multipolarities, and multistep excitations are negligible in
this angular region.
%
On the other hand,
dipole transitions are dominant at forward angles
in Coulomb breakup.
Consequently, as we show in the lower panel, s- and d-wave breakup
cross sections are dominant in this case.
These features also turn out to be the case for breakup reactions by
the other six target nuclei.
Thus, the final states of $^8$B corresponding to
the nuclear breakup and Coulomb
breakup are different from each other,
with respect to the orbital angular momentum
between the $^7$Be and $p$ fragments.
This is the reason for the vanishment of
the N-C interference at $\theta_8 \sim 0^\circ$.
Note that the dominance of the monopole coupling potentials
due to nuclear breakup depends on the $p$-A and
$^7$Be-A potentials. An accurate evaluation of
the distorting potentials used in CDCC calculation is, again, of crucial
importance.

It is sometimes conjectured that the contribution of nuclear breakup
can be quenched if the relative energy $\epsilon$ of the fragments
is truncated at a certain small value $\epsilon_0$.
%
\begin{figure}[htbp]
\begin{center}
\includegraphics[width=0.45\textwidth,clip]{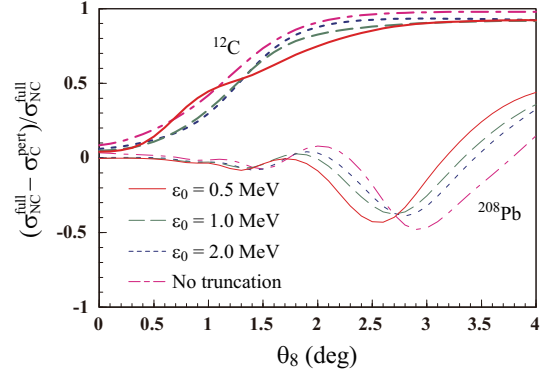}
\end{center}
\caption{
(Color online)
Relative error of the breakup cross section
calculated with first-order perturbative CDCC from
that with full CDCC.
The solid, dashed, and dotted
lines correspond to the results with truncation
of $\epsilon$ at 0.5, 1.0, and 2.0 MeV, respectively.
The results with no truncation are shown by the
dash-dotted lines.
The thick (thin) lines correspond to the breakup
by $^{12}$C ($^{208}$Pb).
}
\label{fig6}
\end{figure}
We show in
Fig.~\ref{fig6} the relative difference between the breakup cross sections
calculated with full CDCC and
first-order perturbative CDCC.
The thick (thin)
lines show the results for the $^{12}$C ($^{208}$Pb) target
with the truncation at $\epsilon_0=0.5$ (solid lines),
1 (dashed lines), and 2 MeV (dotted lines).
The results with no truncation are shown by dash-dotted lines
for comparison.
One sees that the truncation of $\epsilon$
slightly reduces the relative difference at $\theta_8 \sim 0^\circ$.
For $\theta_8 \ga 0.5^\circ$
($2.0^\circ$) for $^{12}$C ($^{208}$Pb), however,
the difference is still very large, which indicates that
even if truncation of $\epsilon$ is carried out,
the first-order perturbative calculation will bring about serious errors
in analysis of the breakup reactions in this angular region,
and hence in the values of the extracted B(E1) and spectroscopic factor.

\section{Summary}
\label{sum}

We reinvestigate the nuclear and Coulomb breakup properties
of $^8$B at 65 MeV/nucleon by the continuum-discretized
coupled-channels method (CDCC) with more reliable $p$-target
and $^7$Be-target optical potentials than in the foregoing work.\cite{Hussein}
The $A_{\rm T}^{1/3}$ scaling law
of the nuclear breakup cross section $\sigma_{\rm N}$,
with $A_{\rm T}$ as the target mass number,
is found to be satisfied only in the middle-mass region, i.e.,
$40 \la A_{\rm T} \la 152$. The interference between nuclear and Coulomb
breakup amplitudes, that is, N-C interference, is very important
even at forward angles, $1^\circ \la \theta_8 \la 2^\circ$,
where $\theta_8$ is the scattering angle of the center of mass of $^8$B.
For the breakup by $^{208}$Pb, it is found that
contributions of $\sigma_{\rm N}$ and N-C interference are not
so important, i.e., less than 10\%, for $\theta_8 \la 2.4^\circ$.
The present calculation
shows that N-C interference indeed tends to vanish
at $\theta_8 \sim 0^\circ$, independently of the target nucleus,
in contrast to the conclusion in
ref.~\citen{Hussein}. Another clarified fact is that the truncation
of the relative energy between the $p$ and $^7$Be fragments
on the measured breakup cross sections quenches the error of the
first-order perturbation theory at $\theta_8 \sim 0^\circ$ but
is not helpful at all for $\theta_8 \ga 0.5^\circ$ ($2.0^\circ$)
in the breakup by $^{12}$C ($^{208}$Pb).
One may conclude from the present study
that the B(E1) strength of $^8$B can be extracted
with quite a small error of about 10\%
from the analysis of the breakup cross section by $^{208}$Pb
for $\theta_8 \la 2.4^\circ$ using
the first-order perturbation theory.
Nevertheless, an assessment of the error of
the perturbation theory is extremely
important because it indeed depends on the reaction systems and
incident energies.
Our final remark is that
all the results
obtained in the present work
reflect the properties of
the optical potentials used in the CDCC calculation.
For systematic analysis
of the breakup reactions of $^{11}$Be, $^{7}$Be, $^{6}$He, $^{11}$Li etc.
with CDCC, an accurate evaluation of the optical potentials concerned
is necessary.

KO thanks S. Bishop for providing elastic cross section data
for $^7$Be-$^{208}$Pb scattering at 60.8 MeV/nucleon.
The computation was mainly carried out using the computer facilities at
the Research Institute for Information Technology, Kyushu University.

\end{document}